\documentclass[12pt,aps,prb,preprint]{revtex4}   

\usepackage{graphicx}   

\newcommand{\mb}[1]{ {{\mathbf{#1}}}}
\newcommand{\of}[1]{\left( #1 \right)}
\newcommand{\grad}{\nabla}
\newcommand{\refeq}[1]{(\ref{#1})}
\newcommand{\half}{\frac{1}{2}}
\newcommand{\Div}[1]{\grad \cdot#1}

\begin{document}
\begin{flushright}
\framebox{\small BRX-TH~547}\\
\end{flushright}

\bibliographystyle{hunsrt}
\title{Schwarzschild and Birkhoff a la Weyl}
\author{S. Deser}
  \affiliation{Department of Physics, Brandeis University, Waltham, MA 02454, USA.}
  \email{deser@brandeis.edu}
\author{J. Franklin}
  \affiliation{Center for Space Research, MIT, Cambridge, MA 02139, USA.}
  \email{jfrankli@mit.edu}


\begin{abstract}
We provide a simple derivation of the Schwarzschild solution in
General Relativity, generalizing an early approach by Weyl, to
include Birkhoff's theorem: constancy of the mass; its deeper,
Hamiltonian, basis is also given. Our procedure is illustrated by
a parallel derivation of the Coulomb field and constancy of
electric charge, in electrodynamics.
\end{abstract}
\maketitle


\section{Introduction}

Deriving and understanding some of the basic properties of the
fundamental -- Schwarzschild -- geometry is a significant hurdle
in elementary expositions of General Relativity (GR). Indeed, no
less distinguished an investigator than Hermann Weyl proudly
discovered an enticing shortcut \cite{Weyl:1993kh}, totally
unjustified at the time, but legitimized much later~\cite{Palais,
Fels:2001rv}.  More recently, it has been used for more
complicated gravity models as well~\cite{DT}.  Weyl's result is,
however, incomplete:  He got the famous $\left(1 - \frac{2 m}{r}
\right)$ factor but assumed {\it a priori}, rather than derived,
the constancy of $m$.  The latter is almost as important a
property as the factor itself, and of course, a consequence of
Einstein's equations.  This property is Birkhoff's
theorem~\cite{BHOFF} -- absence of monopole radiation in GR.  Our
aim here is to retain the attractiveness of Weyl's shortcut, while
simultaneously proving the absence of the $\dot m \ne 0$
``non-solutions". In order to clarify the physics of this
approach, we first establish it in the simpler, but quite
relevant, context of the Coulomb field in electrodynamics.  We
will also briefly discuss the theorem's basis in the deeper
context of the theories' Hamiltonian forms.

\section{Electrodynamics a la Weyl}

We derive the Coulomb field in Maxwell theory, in order to
introduce and suitably extend the Weyl method to include the
vector Birkhoff's theorem -- constancy of electric charge.  Weyl's
general approach was to exploit the special symmetries of the
desired solution by using suitable coordinates and gauges, then
insert the simplified field variables into the action, and vary
only these remaining functions instead of the original set of
variables.

Spherical symmetry means that $\mb{r}$ is the only vector, hence
the $(\mathbf{A}, A_0)$ are restricted to the form
\begin{eqnarray}
\mb{A} = A_r (r,t) \hat\mb{r}  & \quad & A_0 = A_0(r,t) \nonumber \\
\mb{B} = 0 & & \mb{E} =  \of{ A_0' - \dot A_r } \hat\mb{r}
\label{ansatz},
\end{eqnarray}
primes and overdots respectively indicate radial and temporal
derivatives. The vector potential is necessarily a pure gauge, so
can be removed.  As we shall see, this seemingly attractive step
loses the Birkhoff part of Maxwell's equations and hence requires
the additional assumption of time-independence, thereby missing
the fact that the latter is implied by the theory.

Let us now insert \refeq{ansatz} into the Maxwell action,
\begin{equation}\label{maxaction}
I_{Max} = \half \int d^4 x \of{\mb{E}^2 - \mb{B}^2} \; ,
\end{equation}
to obtain the reduced form
\begin{equation}
I_{Max} \rightarrow  2 \pi \int \of{\dot A_r - A_0'}^2 r^2 dr dt;
\end{equation}
we consider only source-free regions throughout. [This approach is
valid in arbitrary dimensions, with $r^2 \rightarrow r^{D-2}$.] If
we impose Coulomb gauge, $A_r = 0$ {\it before} varying, we
immediately obtain the single field equation $\of{r^2 A_0'} ' =
0$, whose solution is of course $A_0 = q(t)/r$.  However, we
cannot then infer $\dot q = 0$, the subset that solves Maxwell's
equations.  If instead, we only set $A_r = 0$ after varying
\refeq{maxaction}, we learn that the time derivative of the field
equation also vanishes -- the variation of the gauge part gives
\begin{equation}\label{Lfield}
\frac{\delta I}{\delta A_r}\biggr\vert_{A_r = 0} =  4\pi r^2 \dot
A_0'(r,t) = 0 = \dot q
\end{equation}
Of course, we need not set $A_r = 0$ at all, the gauge invariant
content of \refeq{maxaction} being
\begin{equation}
\Div{\mb{E}} = 0 = \Div{\dot\mb{E}},
\end{equation}
since varying $A_r$ and $A_0$ manifestly yields the respective
time and space derivatives of the same quantity, namely $\mb{E}$.
Note that the second equation
\begin{equation}\label{EMBianchi}
\Div{\dot \mb{E}} = \Div{\of{\grad\times \mb{B}}} = 0,
\end{equation}
reflects the Bianchi identities $\partial_\mu \of{\partial_\nu
F^{\mu \nu}} = 0$, i.e. $\partial_0 \of{\partial_i F^{0 i}} +
\partial_j \of{\partial_\mu F^{\mu j}} = 0$, as the last term is a
field equation.

The time-constancy of spherical solutions is manifest in the
action's Hamiltonian form, where $(-\mb{E},\mb{A})$ are
independent variables, the canonical ``$(p,q)$" pairs:
\begin{equation}
I_{Max} = -\int d^4 x \left[ \mb{E}^T \cdot \dot\mb{A}^T +
\mb{E}^L \of{ \dot\mb{A}^L - \mbox{\boldmath$\nabla$} A_0 } +
\half \left\{  \mb{E}^2 + \of{\mbox{\boldmath$\nabla$} \times
\mb{A}^T}^2\right\}\right].
\end{equation}
We have used the orthogonal decomposition of a vector,
\begin{equation}
\mb{V} = \mb{V}^T + \mb{V}^L \; , \;\;\; \Div \mb{V}^T \equiv 0
\equiv \nabla \times \mb{V}^L \; , \;\;\; \int d^3r \: V^T \cdot
W^L = 0 \; .
\end{equation}
 Since time-dependence only appears in the ``$p \dot q$" terms, and
there are no transverse spherically symmetric vectors, we learn
immediately from varying the surviving component, $\mb{A}^L$, that
$\dot\mb{E}^L = 0$, the rest of the action being
$\mb{A}^L$-independent.

The lesson, one that will carry over unaltered to GR, is that
Weyl's approach, using as few functions as gauge choice allows,
lulls one into the unjustified belief that all is time-independent
just because only spatial derivatives remain. This point is
relevant because, even if one does not assume time-independence
but prematurely drops $A_r$, the resulting equation for $A_0$ has
no explicit time-derivatives.

This is a good place to discuss the validity of the Weyl procedure
itself. For the linear Maxwell theory, it is easy enough to
understand the ``commutativity" between first inserting a
symmetric ansatz in an action before varying, or only doing so
after full variation.  Clearly, $\dot \mb{E} = \grad \times
\mb{B}$ and $\Div{\mb{E}} = 0$ immediately degenerate, with the
spherical symmetry requirement that $\mb{E} = \grad{\phi}$,
$\mb{B} = 0$, into $\nabla^2 \phi = 0$, $\dot \phi = 0$, which can
then be variously decomposed in different gauges.  So no
information is lost by varying the action if (and only if) both
functions are kept.

More generally, it is intuitively pretty clear that, since the
solutions are extrema also within the set of spherically symmetric
trial variables, we will not get any false ones.  That we will
also not miss any true solutions in this way is pretty reasonable
as well.  We will, however, say no more on this deep and difficult
problem nor on its extensive fine print;  for this, we refer to
the original work~\cite{Palais} and to a later exegesis
specifically in GR~\cite{Fels:2001rv}.  Some of the perils
involved are illustrated in a recent note~\cite{DFT}.

\section{GR Weyl -- Static}
The stage has now been set for our GR target.  We approach it in
two steps. The first still adheres to the original Weyl line,
losing time-independence.  That will be followed by the full
Birkhoff treatment.

We begin with the general form of a spherically symmetric metric
tensor $g_{\mu \nu}$ or its corresponding interval $ds^2 = g_{\mu
\nu} dx^\mu dx^\nu$.  For completeness, we first set to rest the
misplaced worry that spherical symmetry cannot mean anything in a
theory with general coordinate invariance; this is a
misunderstanding of coordinates, having nothing to do with
geometry.  The fancy answer is that symmetries are characterized
by the existence of (one or more) Killing vectors $X_\mu^{(a)}$
obeying the invariant equation $D_\mu X_\nu^{(a)} + D_\nu
X_\mu^{(a)} \equiv \partial_\mu X^{(a)}_\nu + \partial_\nu
X^{(a)}_\mu - (\partial_\nu g_{\mu\alpha} + \partial_\mu
g_{\nu\alpha} - \partial_\alpha g_{\mu\nu}) X^{\alpha(a)} = 0$.
For example, if there is an $X_\mu^{(a)}$ which takes the
non-invariant form $X_\mu^{(a)} = g_{\mu a}$ in some coordinates,
then the Killing equation immediately reduces to the statement
that (in that frame) $\partial_a g_{\mu\nu}$ = 0.  But it suffices
simply to remember what spherical symmetry means in Cartesian
coordinates:  given that $x^i$ is the only vector and
$\delta_{ij}$ the only $2$-tensor that can appear, then
\begin{equation}
g_{ij} = A \delta_{ij} + B x^i x^j \quad g_{0i} = C x^i \quad
g_{00} = - D
\end{equation}
where $(A,B,C,D)$ depend only on $r^2 \equiv \of{x^2 + y^2 + z^2}$
and any other ``irrelevant" parameters such as time.  The
corresponding interval is:
\begin{equation}\label{dsgeneric}
ds^2 = -D dt^2 + (A + Br^2) dr^2 + Ar^2 d\Omega + 2 Cr dr dt
\end{equation}
where $d\Omega$ is the usual unit $2$-sphere element, since $x^i
x^j dx^i dx^j \equiv r^2 dr^2$, $\delta_{ij} dx^i dx^j = dr^2 +
r^2 d\Omega$.  This four-function parametrization really consists
of two physical, plus two gauge, components -- double the
$(A_0,A_r)$ set of vector theory.  Weyl's choice was to
diagonalize away the $dr dt$ term, and use Schwarzschild
coordinates, $A = 1$, leaving just one spatial and one temporal
metric component.  We begin with the more instructive choice in
which all three functions $(A, B, D)$ are kept, but still dropping
the off-diagonal $C$.  The latter is in fact precisely the analog
of the first pass in Maxwell theory, so we will not yet achieve
Birkhoff's theorem; indeed, this pinpoints where the original Weyl
ansatz is insufficient and requires the redundant assumption of
time-independence.

Our starting point then is the $3$-function interval
\begin{equation}
ds^2 = -a b^2 dt^2 +a^{-1} dr^2 + c^2 d\Omega
\end{equation}
where we have made things a lot easier to calculate by the above
$(a,b)$ parametrization.  There is no loss of generality in this,
just looking ahead to the $b = 1$ result for Schwarzschild.

Calculation of the curvature cannot be avoided, even here, but it
is mercifully short and yields
\begin{eqnarray}
I_E &= &\int d^4 x \sqrt{-g} R \Rightarrow I_E(a,b,c) = I_r + I_t \\
I_r &=& 8 \pi \int dt dr \of{a b' (c^2)' + b \of{1 + c' \of{a c}'}} \\
I_t &=& 8\pi \int dt dr \frac{\dot c}{a^2 b} \of{c \dot a - a \dot c}.
\end{eqnarray}
The two parts $I_r$ and $I_t$ of the action contain either space
or time derivatives, but not both.  The original $2$-function Weyl
ansatz was to set $c = r$ {\it before} varying, which is why he
would never see $I_t$ but only $I_r$;  this leaves
\begin{equation}
I_W(a,b) = 8\pi \int dr \of{b + r a b'}
\end{equation}
which immediately yields the ``Schwarzschild" result $a = 1 -
\frac{2 m}{r}$, and $b = b_0$, but with possibly time-dependent
$(b_0, m)$. Time-dependence of $b_0$ is irrelevant as it can be
absorbed into $dt$ by fixing the remaining $t \rightarrow t' (t)$
gauge freedom.

If we keep all three functions, but drop the time-dependence, then
$(a,b)$ are parametrized by $c$, which stays undetermined:
 \begin{eqnarray}
a&= &\frac{1}{c'^2} \of{1 - \frac{2 m}{c}} \\
b &=& b_0 c'  \nonumber,
 \end{eqnarray}
 corresponding to the interval
 \begin{equation}\label{Schwarzschildds}
 ds^2 = -b_0^2 \of{1 - \frac{2 m}{c}}dt^2 + \frac{1}{1  -\frac{2 m}{c}} dc^2 + c^2 d\Omega,
  \end{equation}
using $dr^2 = dc^2/c'^2$.  This result shows the very special role
played by Schwarzschild coordinates; they are not so much a gauge
as the natural parametrization of the interval in terms of the
$2$-sphere ``orbits".  Indeed, writing $c = r$ in
\refeq{Schwarzschildds} is more an exercise in penmanship than a
choice of gauge!

\section{Birkhoff's Theorem}

The Maxwell example linked absence of monopole radiation to that
of ``scalar" -- helicity zero -- modes.  Let us first turn to
linearized gravity, its direct counterpart.  Here the Hamiltonian
form is expressed in terms of the conjugate pair of spatial
tensors $(\pi^{i j}, \linebreak h_{ij} \equiv g_{ij} -
\delta_{ij})$. The tensorial orthogonal transverse-longitudinal
decomposition can be written as
\begin{equation}
h_{ij} = h_{ij}^{TT} + \of{\partial_ih^T_j + \partial_j h_i^T} +
\nabla^{-2} \partial^2_{ij}h^L + \half\of{\partial^2_{ij} -
\frac{\delta_{ij}}{\nabla^2} }h^T.
\end{equation}
For our purposes, it suffices to note that spherically symmetric
tensors lack the transverse-traceless (TT) tensor quadrupole, and
transverse vector ($h_i^T$) dipole, modes.  The action
$$\label{linGR}
 I_{\ell in}[\pi, h] = \int d^3 r dt [ \pi^{ij}h_{ij} - H (\pi , h)]
 \eqno{(19{\rm a})}
 $$
then reduces to~\cite{ADM}
 $$
 I_{\ell in}[\pi, h] \rightarrow \int d^3 r dt \of{\pi^L \dot{h}^L
  + \half \pi^T \dot h^T - H}. \eqno{(19{\rm b})}
 $$
\setcounter{equation}{19}
 The Hamiltonian's details are irrelevant:
all that counts for us is its (abelian) gauge invariance, that it
is independent of the two gauge functions $(h^L, \pi^T)$. Hence,
just as in Maxwell theory, we may immediately conclude from their
variation that
 \begin{equation}
 \partial_0 h^T = 0 = \partial_0 \pi^L  .
 \end{equation}
This time-independence, equivalent to $\partial_0 \of{\grad\cdot
\mb{E}} = 0$ is also a direct consequence of the linearized
Bianchi identities which state that (on shell) $\partial_0 G_{\ell
in} ^{0\mu} = 0$, precisely the same two statements as
\refeq{EMBianchi}.  Since our fields are tensorial, they have four
(energy-momentum) conservation laws rather than the single one of
electrodynamics, though here there is only radial momentum left.
The full GR action can also be gauge-fixed to the simple
(seeming!) form\cite{ADM}.
 \begin{equation}
 I_E = \int d^4 x \of{\pi^{ij}_{TT} \dot g^{TT}_{ij} - H(TT)} \; .
  \end{equation}
The only time dependence is in the ``TT" modes.  Thus Birkhoff's
theorem  holds also in full GR and indeed even rids one of dipole
radiation since dipoles cannot construct ``TT" tensors  -- in
either the linearized or the full theory.

Let us now proceed to our concrete Weyl setting.  Instead of
keeping all four metric components in the generic interval
\refeq{dsgeneric}, we just introduce the off-diagonal one, which
is effectively the above gauge function $\pi^T$ in the spherical
case. In order to avoid pedantic overkill, let us use
Schwarzschild coordinates {\it ab initio} here (since $c$ just
defines the ``radial" coordinate, we lose nothing by doing so from
the start) and concentrate on the three $(r,t)$-dependent
functions $(a,b,f)$
 \begin{equation}
 ds^2 = -a b^2 dt^2 + a^{-1} dr^2 + r^2 d\Omega + 2 b f dr dt;
 \end{equation}
writing the cross term as $b f$ simplifies the calculation. The
full three-function action is neither pretty nor useful:  all we
need are the new terms linear in $f$, since we will set $f = 0$
after varying anyway (analogous to the electrodynamics case, where
the gauge $A_r = 0$ is our choice).  The action then simply
reduces to the old Weyl term $I(a,b)$ plus the $f$-term that will
act as a Lagrange multiplier enforcing time-independence of $m$ as
follows:
\begin{equation}
I(a,b,t) = 8\pi \int dr dt \of{b \of{r - a r}' + a^{-1} r f \dot
a}.
\end{equation}
The respective variations then yield (at $f = 0$) the desired
constant mass geometry:
 \begin{eqnarray}
 \frac{\delta I}{\delta a}\biggr\vert_{f = 0}  &=  & r b' = 0  \Rightarrow b = b(t)  \\
 \frac{\delta I}{\delta b}\biggr\vert_{f = 0}  &= & \of{r - a r}' = 0  \Rightarrow a = 1 - \frac{2 m(t)}{r} \\
 \frac{\delta I}{\delta f}\biggr\vert_{f = 0}  &= & ra^{-1}\dot a = 0  \Rightarrow \dot m = 0.
 \end{eqnarray}
As before, the time-dependence of $b_0(t)$ can be removed by a
pure time redefinition, leaving us with the correct Birkhoff
statement $\dot m = 0$ as the gauge-varied field equation, just as
$\dot q = 0$ came from the analogous one in electrodynamics.

\section{Summary}
We have attempted to present a logical and intuitive basis for
deriving and understanding the Schwarzschild solution and its
time-independence.  We followed the Maxwell example to display the
pattern of ``true-plus-gauge" variables and the role played by the
latter in ensuring constancy of the corresponding ``charges".

Using the suitably extended Weyl method enabled us to avoid as
much tensorial machinery as possible while still keeping all the
implications of Einstein's equations.  One obvious future
application is to the considerably more complicated Kerr solution,
the time-independent but rotating (stationary) dipole metric.
Indeed, a useful exercise for the interested student would be to
derive the linearized version of this geometry!

\section*{Acknowledgements}
We variously thank B.\ Tekin, I.B.\ Khriplovich, D.\ Schroeter and
a very conscientious referee for demanding (if not obtaining) more
clarity. S. Antoci and I. Bengtsson provided historical
background. This work was supported by NSF grant PHY04-01667.

\bibliography{alaweyl03.bbl}
 \end{document}